# Low Power Reversible Parallel Binary Adder/Subtractor

Rangaraju H G[1], Venugopal U[2], Muralidhara K N[3], Raja K B [2]

[1]Department of Telecommunication Engineering, Bangalore Institute of Technology, Bangalore, India
rang_raju@yahoo.com

[2]Department of Electronics and Communication Engineering, University Visvesvaraya College of Engineering, Bangalore, India
venu.ubaradka@gmail.com

[3]Department of Electronics and Communication Engineering, P E S College of Engineering, Mandya, Karnataka, India

## Abstract

*In recent years, Reversible Logic is becoming more and more prominent technology having its applications in Low Power CMOS, Quantum Computing, Nanotechnology, and Optical Computing. Reversibility plays an important role when energy efficient computations are considered. In this paper, Reversible eight-bit Parallel Binary Adder/Subtractor with Design I, Design II and Design III are proposed. In all the three design approaches, the full Adder and Subtractors are realized in a single unit as compared to only full Subtractor in the existing design. The performance analysis is verified using number reversible gates, Garbage input/outputs and Quantum Cost. It is observed that Reversible eight-bit Parallel Binary Adder/Subtractor with Design III is efficient compared to Design I, Design II and existing design.*

## 1. Introduction

Reversible computing was started when the basis of thermodynamics of information processing was shown that conventional irreversible circuits unavoidably generate heat because of losses of information during the computation [1]. The different physical phenomena can be exploited to construct reversible circuits avoiding the energy losses. One of the most attractive architecture requirements is to build energy lossless small and fast quantum computers. Most of the gates used in digital design are not reversible for example NAND, OR and EXOR gates.

A Reversible circuit/gate can generate unique output vector from each input vector, and vice versa, i.e., there is a one to one correspondence between the input and output vectors. Thus, the number of outputs in a reversible gate or circuit has the same as the number of inputs, and commonly used traditional NOT gate is the only reversible gate. Each Reversible gate has a cost associated with it called Quantum cost. The Quantum cost of a Reversible gate is the number of 2*2 Reversible gates or Quantum logic gates required in designing. One of the most important features of a Reversible gate is its garbage output i.e., every input of the gate which is not used as input to other gate or as a primary output is called garbage output.

In digital design energy loss is considered as an important performance parameter. Part of the energy dissipation is related to non-ideality of switches and materials. Higher levels of integration and new fabrication processes have dramatically reduced the heat loss over the last decades. The power dissipation in a circuit can be reduced by the use of Reversible logic. Landauer's [2] principle states that irreversible computations generates heat of K*Tln2 for every bit of information lost, where *K* is Boltzmann's constant and T the absolute temperature at which the computation performed. Bennett [3] showed that if a computation is carried out in Reversible logic zero energy dissipation is possible, as the amount of energy dissipated in a system is directly related to the number of bits erased during computation. The design that does not result in information loss is irreversible. A set of reversible gates are needed to design reversible circuit. Several such gates are proposed over the past decades.

Arithmetic circuits such as Adders, Subtractors, Multipliers and Dividers are the essential blocks of a Computing system. Dedicated Adder/Subtractor circuits are required in a number of Digital Signal Processing applications. Several designs for binary Adders and Subtractors are investigated based on Reversible logic. Minimization of the number of Reversible gates, Quantum cost and garbage inputs/outputs are the focus of research in Reversible logic.

*Contribution:* In this paper, novel three Design types viz., Design I, Design II and Design III of Reversible Eight-bit Parallel Binary Adder/Subtractor are proposed. The Reversible gates such as F, FG, TR and PG are used to construct Design I, Design II and Design III Adder/Subtractor. The performance of Design III is better in terms of number of gates, Garbage inputs/outputs and Quantum Cost in comparison with Design I and Design II.

*Organization:* The paper is organized into the following sections. Section 2 is an overview of Reversible gates. The Background work is described in section 3. Section 4 is the proposed design, Result analysis of the proposed design is presented in section 5 and Conclusions are contained in section 6.

## 2. Reversible Gates

The simplest Reversible gate is NOT gate and is a 1*1 gate. Controlled NOT (CNOT) gate is an example for a 2*2 gate. There are many 3*3 Reversible gates such as F, TG, PG and TR gate. The Quantum Cost of 1*1 Reversible gates is zero, and Quantum Cost of 2*2 Reversible gates is one. Any Reversible gate is realized by using 1*1 NOT gates and 2*2 Reversible gates, such as V, $V^+$ (V is square root of NOT gate and $V^+$ is its hermitian) and FG gate which is also known as CNOT gate. The V and $V^+$ Quantum gates have the property given in the Equations 1, 2 and 3.

$$V * V = NOT \quad \ldots\ldots\ldots\ldots\ldots (1)$$

$$V * V^+ = V^+ * V = I \quad \ldots\ldots\ldots\ldots (2)$$

$$V^+ * V^+ = NOT \quad \ldots\ldots\ldots\ldots\ldots (3)$$

The Quantum Cost of a Reversible gate is calculated by counting the number of V, $V^+$ and CNOT gates.

### 2.1 NOT Gate

The Reversible 1*1 gate is NOT Gate with zero Quantum Cost is as shown in the Figure 1.

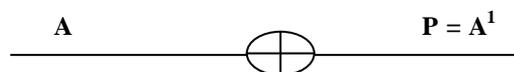

Figure1. NOT gate

### 2.2 Feynman / CNOT Gate

The Reversible 2*2 gate with Quantum Cost of one having mapping input (A, B) to output (P = A, Q = A ⊕ B) is as shown in the Figure 2.

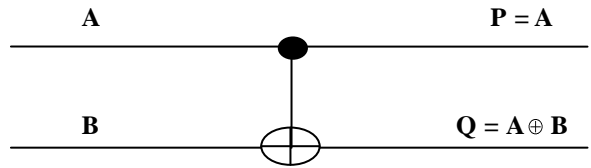

Figure2. Feynman/CNOT gate

## 2.3 Toffoli Gate

The 3*3 Reversible gate with three inputs and three outputs. The inputs (A, B, C) mapped to the outputs (P=A, Q=B, R=A.B $\oplus$ C) is as shown in the Figure 3.

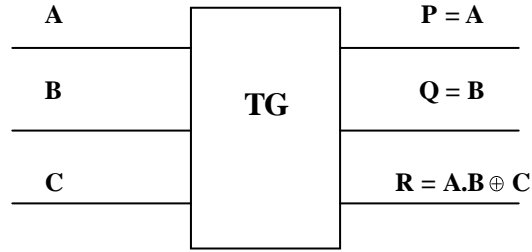

Figure3. Toffoli gate

Toffoli gate [4] is one of the most popular Reversible gates and has Quantum Cost of 5. It requires 2V, 1 V$^+$ and 2 CNOT gates. Its Quantum implementation is as shown in Figure 4.

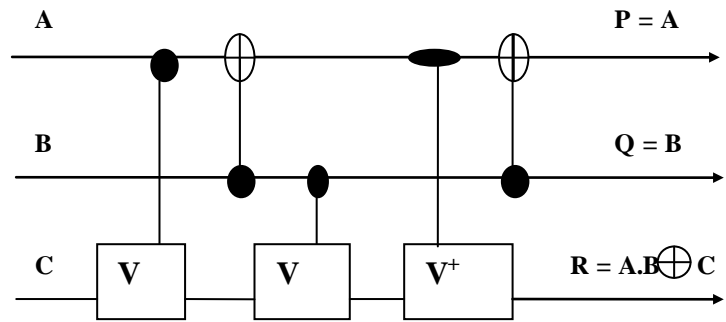

Figure4. Quantum implementation of Toffoli gate

## 2.4 Peres Gate

The three inputs and three outputs i.e., 3*3 reversible gate having inputs (A, B, C) mapping to outputs (P = A, Q = A $\oplus$ B, R = (A.B) $\oplus$ C). Since it requires 2 V$^+$, 1 V and 1 CNOT gate, it has the Quantum cost of 4. The Peres gate and its Quantum implementation are as shown in the Figure 5 and 6 respectively.

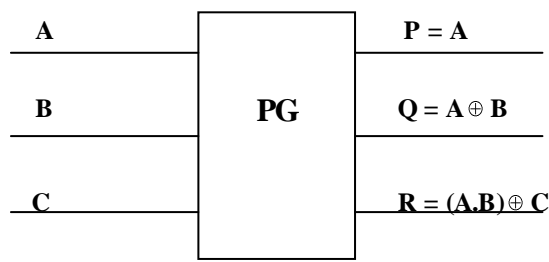

Figure5. Peres gate

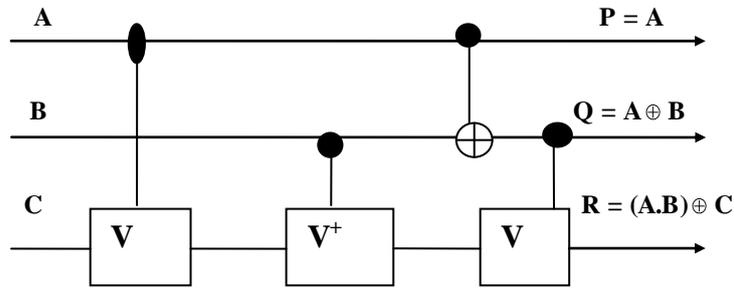

Figure6. Quantum implementation of Peres gate

## 2.5 Fredkin Gate

Reversible 3*3 gate maps inputs (A, B, C) to outputs (P=A, Q=A'B+AC, R=AB+A'C) having Quantum cost of 5 and it requires two dotted rectangles, is equivalent to a 2*2 Feynman gate with Quantum cost of each dotted rectangle is 1, 1 V and 2 CNOT gates. Fredkin gate and its Quantum implementations are shown in Figure 7 and 8 respectively.

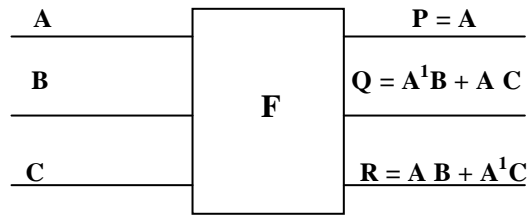

Figure7. Fredkin gate

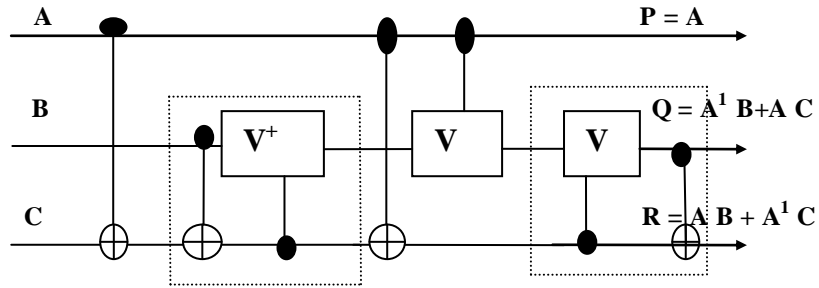

Figure8. Quantum implementation of Fredkin gate

## 2.6 TR Gate

The gate has 3 inputs and 3 outputs having inputs (A, B, C) mapped to the outputs (P=A, Q=A $\oplus$ B, R= (A.B$^1$) $\oplus$ C). TR gate is shown in Figure 9.

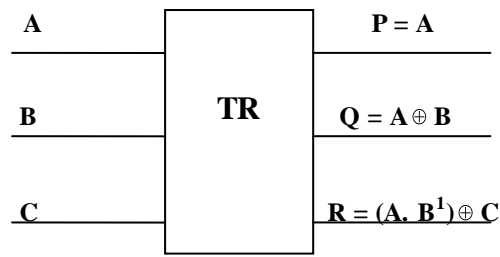

Figure9. TR gate

The Quantum cost of TR gate can be estimated by realizing from 1 Toffoli gate, 2 NOT gates and 1 CNOT gate as shown in the Figure 10. Thus the Quantum cost of TR gate will be Quantum cost of CNOT gate plus Quantum cost of 1 Toffoli gate which is equal to 6.

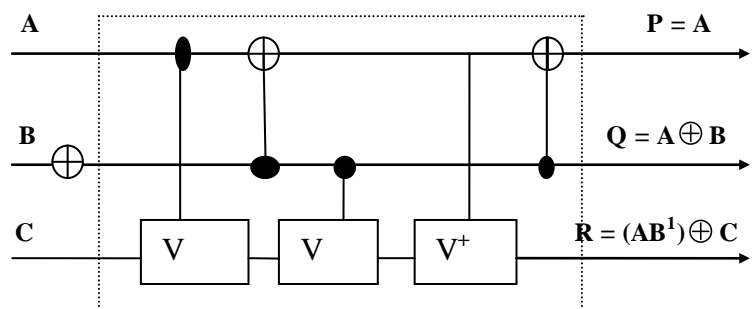

Figure 10. Quantum implementation of TR gate

# 3. Literature Survey

Thapliyal and Ranganathan [5] proposed the design of Reversible Binary Subtractor using TR Gate. The particular function like Binary Subtraction is implemented using TR gate effectively by reducing number of Reversible gates, Garbage outputs and Quantum Cost. Thapliyal and Ranganathan [6] presented a design of Reversible latches viz., D Latch, JK latch, T latch and SR latch that are optimized in terms of quantum cost, delay and garbage outputs.. Lihui Ni et al., [7] described general approach to construct the Reversible full adder and can be extended to a variety of Reversible full-adders with only two Reversible gates. Irina Hashmi and Hafiz Hasan Babu [8] designed an efficient reversible barrel shifter which is capable of left shift/rotate used for high speed ALU applications. Robert Wille et al., [9] explored two techniques from irreversible equivalence checking applied in the reversible circuit domain. (i) Decision diagram Technique equivalence checking for quantum circuits and (ii) Boolean satifiability checking for garbage input/outputs. Noor Muhammed Nayeem et al., [10] presented designs of Reversible shift registers such as serial-in serial-out, serial-in parallel-out, parallel-in serial-out, parallel-in parallel-out and universal shift registers. Majid Mohammadi, Mohammad Eshghi et al., [11] proposed a synthesis method to realize a Reversible Binary Coded Decimal adder/subtractor circuit. Genetic algorithms and don't care concepts used to design and optimize all parts of a Binary Coded Decimal adder circuit in terms of number of garbage inputs/outputs and quantum cost.

Majid Mohammadi and Mohammad Eshghi [12] explained about the behavioral description and synthesis of quantum gates. To synthesize reversible logic circuits, V and V+ gates are shown in the truth table form and shown that bigger circuits with more number of gates can be synthesized. Rekha James et al., [13] proposed an implementation of Binary Coded Decimal adder in Reversible logic, which is basis of ALU for reversible CPU. VLSI implementations using one type of building block can decrease system design and manufacturing cost. Himanshu Thapliyal and Vinod [14] presented the Transistor realization of a new 4*4 Reversible TSG gate. The gate alone operates as a Reversible full adder. The Transistor realizations of 1-bit Reversible full adder, ripple carry adder and carry skip adder are also discussed. Himanshu Thapliyal and Srinivas [15] proposed a 3x3 Reversible TKS gate with two of its outputs working as 2:1 multiplexer. The gate used to design a Reversible half adder and further used to design multiplexer based Reversible full adder. The multiplexer based full adder is further used to design Reversible 4x4 Array and modified Baugh Woolley multipliers.

Yvan Van Rentergem and Alexis De Vos [16] presented four designs for Reversible full-adder circuits and the implementation of these logic circuits into electronic circuitry based on C-MOS technology and pass-transistor design. The chip containing three different Reversible full adders are discussed. Mozammel Khan [17] proposed realizations of ternary half and full-adder circuits using generalized ternary gates. Mozammel Khan [18] discussed quantum realization of ternary Toffoli gate which requires fewer gates than the existing literature. Abhinav Agrawal and Niraj Jha [19] presented the first practical synthesis algorithm and tool for Reversible functions with a large number of inputs. It uses positive-polarity Reed-Muller decomposition at each stage to synthesize the function as a network of Toffoli gates. Pawel Kerntopf [20] explained multipurpose Reversible gates and example of efficient binary multipurpose reversible gates.

## 4. Proposed Model

### 4.1 Adder/Subtractor – Design I

#### 4.1.1 Half Adder/Subtractor

Reversible half Adder/Subtractor–Design I is implemented with four Reversible gates of which two F and two FG gates is shown in the Figure 11. The numbers of Garbage outputs are three represented as $g_1$ to $g_3$, Garbage inputs are two represented by logical *zero* and Quantum Cost is twelve as it requires two FG gates each costing one and two F gates each costing five each.

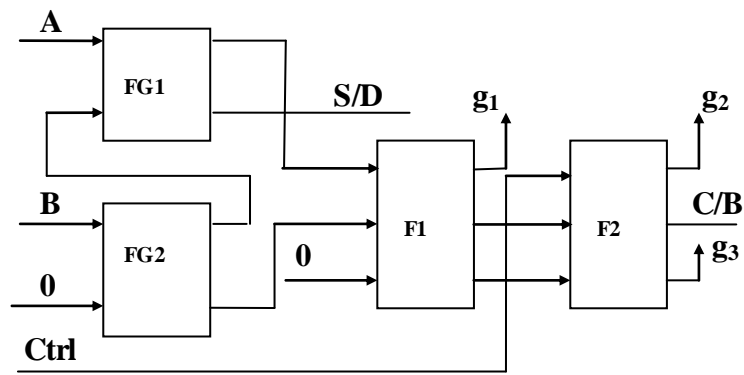

Figure11. Reversible Half Adder/Subtractor – Design I

#### 4.1.2 Full Adder/Subtractor

The Design I Reversible Full Adder/Subtractor with five FG, two F and a TR gate is as shown in the Figure 12. The three inputs are A, B and Cin and the outputs are Sum/Difference (S/D) and Carry/Borrow (C/B). The Control (Ctrl) input differentiates the Addition and Subtraction functionalities. For Ctrl value *zero* i.e., Logical *low* the circuit performs addition and Subtraction for Ctrl value *one* i.e., Logical *high*. The numbers of Garbage inputs are 3 represented by logical *zero*. The Garbage outputs are 5 represented by $g_1$ to $g_5$. The Sum/Difference function is realized from FG4 gate, and the Carry/Borrow function is realized from the output of TR gate. The Quantum Cost for five FG gates are five as each gate costs one, for two F gates is ten as each gate costs five, one TR gate costs six and total design Quantum Cost is 21.

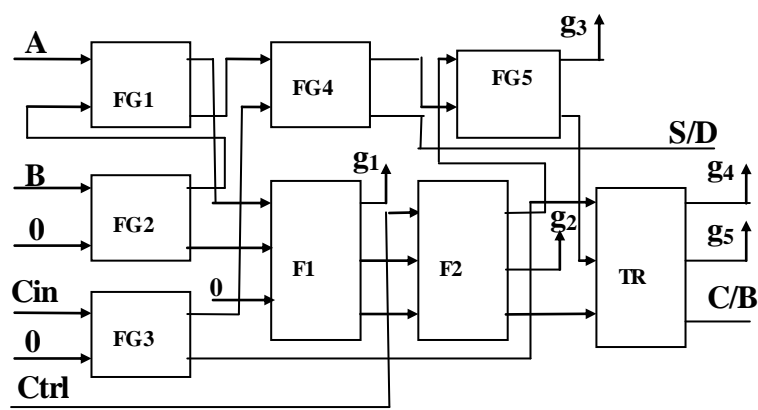

Figure12. Reversible Full Adder/Subtractor – Design I

## 4.2 Adder/Subtractor – Design II

### 4.2.1 Half Adder/Subtractor

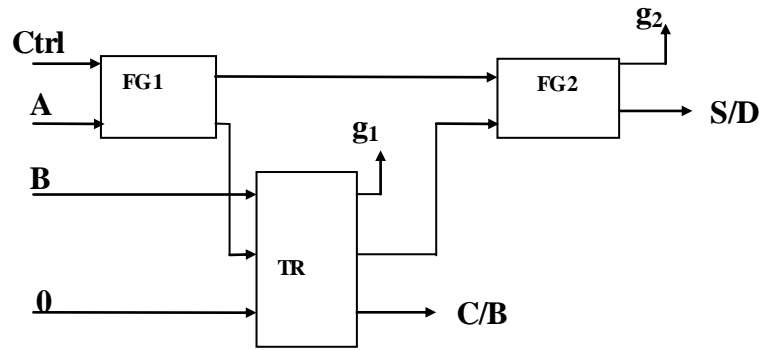

Figure 13. Reversible Half Adder/Subtractor – Design II

Reversible half Adder/Subtractor–Design II is implemented with three Reversible gates of which two are FG gates with each having Quantum cost of one and a TR gate with six Quantum cost is as shown in the Figure 13. The number of Garbage outputs is two i.e., $g_1$ and $g_2$, Garbage inputs one denoted by logical *zero* and total Quantum Cost is eight.

### 4.2.2 Full Adder/Subtractor

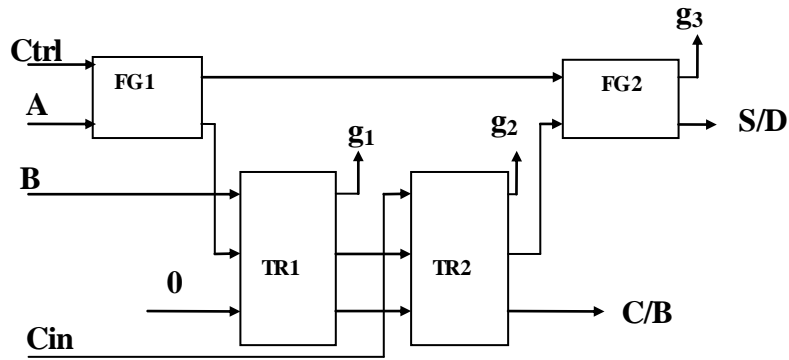

Figure 14. Reversible Full Adder/Subtractor – Design II

Two TR gates and two FG gates are used to realize Deign II Reversible Full Adder/Subtractor unit is shown in Figure 14. The three inputs are A, B and Cin, the outputs are S/D and C/B. For Ctrl value *zero* the circuit performs addition and Subtraction for Ctrl value *one*. The numbers of Garbage inputs are 1 represented by logical *zero*. The Garbage outputs are 3 represented by $g_1$ to $g_3$. The Quantum Cost for the design is 14. A Quantum Cost advantage of 7 is obtained when compared to Adder/Subtractor Design I. Quantum Cost advantage is due to the realization of Arithmetic blocks (Adder and Subtractor) using two TR gates as against three numbers of 3*3 gates for Design I.

## 4.3 Adder/Subtractor – Design III

### 4.3.1 Half Adder/Subtractor

Reversible half Adder/Subtractor–Design III is implemented with three Reversible gates of which two are FG gates each having Quantum cost of one and a PG gate with Quantum cost four is as shown in the Figure 15. The numbers of Garbage outputs is two i.e., $g_1$ and $g_2$, Garbage inputs are one denoted by logical *zero* and Quantum Cost is six.

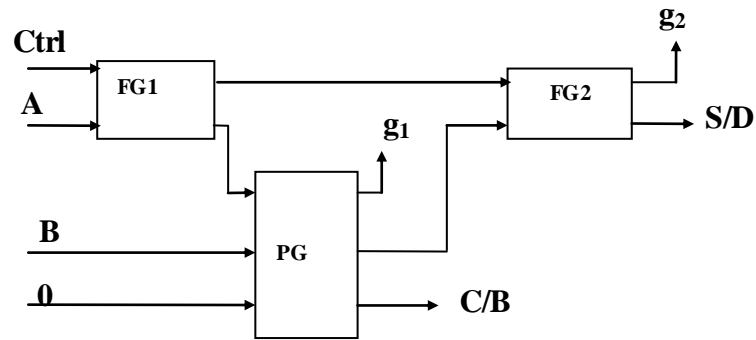

Figure15. Reversible Half Adder/Subtractor – Design III

### 4.3.2 Full Adder/Subtractor

The Reversible Full Adder/Subtractor Design III consists of two FG, two PG gates, and their interconnections are shown in the Figure 16. The three inputs are A, B, and Cin, The outputs are S/D and C/B. For Ctrl value *zero* the circuit performs addition and Subtraction for Ctrl value *one*. The numbers of Garbage inputs are 1 represented by logical *zero*. The Garbage outputs are 3 represented by $g_1$ to $g_3$. The Quantum Cost for the design is 10. A Quantum Cost advantage of 11 is obtained when compared to Adder/Subtractor Design I and of 4 when compared to Adder/Subtractor Design II. Quantum Cost advantage is due to the realization of Arithmetic blocks using two PG gates as against two F and one TR gate for Design I and two TR gates for Design II.

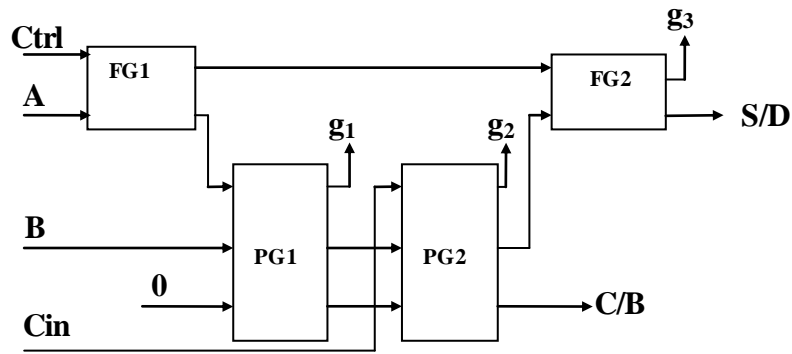

Figure16. Reversible Full Adder/Subtractor – Design III

### 4.4 Reversible Eight-bit Parallel Binary Adder/Subtractor

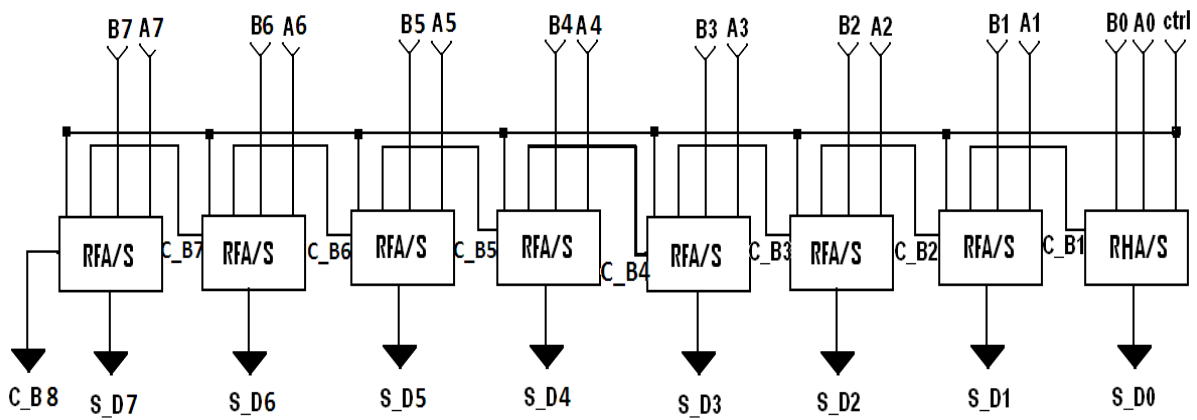

Figure17. Reversible Eight-bit parallel Binary Full Adder/Subtractor

The Half and Full Adder/Subtractor Design I, Design II and Design III are used to construct Reversible eight-bit Parallel Binary Adder/Subtractor is shown in the Figure 17. The ctrl input is used to differentiate eight-bit addition and subtraction functions. The two eight-bit binary numbers are A0 to A7 and B0 to B7. Carry/Borrow is obtained after Addition/Subtraction is represented by C_B1 to C_B7. The outputs Sum/Difference and Carry are shown as S_D0 to S_D7 and C_B8 respectively. The implementation requires seven Full Adder/Subtractor units and one half Adder/Subtractor units in which first stage is half Adder/Subtractor.

# 5. Results

## 5.1 Reversible Full Adder/Subtractor

The comparison of Reversible Full Adder/Subtractor Design I, Design II and Design III in terms of the number gates, number of Garbage inputs/outputs and Quantum Cost of the logics is shown in the Table 1.

It is observed that Design III has better performance compared to Design II and Design I. The number of Reversible gates required for Design III is only 4 as compared to 8 and 4 in the cases of Design I and II respectively, which indicates that the improvement of 100% compared to Design I. The Garbage outputs are 5 in the case of Design I, whereas 3 in the case of Design II and Design III, i.e., the improvement is 65% in Design III compared to Design I. The Garbage inputs are 3 in the case of Design I and one in case of Design II and Design III, gives 200% improvement in Design III compared to Design I. Quantum Cost of Design III, Design II and Design I are 21, 14 and 10 respectively, resulting in improvement of Design III over Design II and Design I are 40% and 110% respectively.

Table1. Comparison of Reversible Full Adder/Subtractor

|  | **Reversible Gates** | **Garbage outputs** | **Garbage inputs** | **Quantum Cost** |
|---|---|---|---|---|
| **Add/Sub–Design I** | **08** | **05** | **03** | **21** |
| **Add/Sub–Design II** | **04** | **03** | **01** | **14** |
| **Add/Sub–Design III** | **04** | **03** | **01** | **10** |

## 5.2 Reversible eight-bit Parallel Binary Adder/Subtractor

The number of gates, Garbage inputs/outputs and Quantum Cost for Reversible eight-bit parallel binary Adder/Subtractor Design I, Design II and Design III are compared as shown in the Table 2.

Table2. Comparison of Reversible eight-bit Parallel Binary Adder/Subtractor design

|  | **Reversible Gates** | **Garbage outputs** | **Garbage inputs** | **Quantum Cost** |
|---|---|---|---|---|
| **Add/Sub– Design I** | **60** | **38** | **23** | **159** |
| **Add/Sub–Design II** | **31** | **23** | **08** | **106** |
| **Add/Sub- Design III** | **31** | **23** | **08** | **76** |

It is seen that Design III has better performance compared to Design II and Design I. The number of Reversible gates required for Design III is 31 as compared to 60 and 31 in the cases of Design I and II respectively, which is an improvement of 93.5% compared to Design I. The Garbage outputs are 38 in the case of Design I, whereas 23 in the case of Design II and Design III, yields an improvement of 65.21% in Design III compared to Design I. The Garbage inputs are 23 in the case of Design I and 8 in case of Design II and Design III, resulting 187.5% improvement in Design III compared to Design I. Quantum Cost of Design III, Design II and Design I are 76, 106 and 159 respectively, hence an improvement of Design III over Design II and Design I are 39.47% and 109.20% respectively.

The existing Reversible Binary Subtractor based on Reversible gate [5] to implement full Subtraction requires Quantum Cost of 12, Garbage inputs of *one* and Garbage outputs of *two*. The proposed Reversible eight-bit Parallel Binary Adder/Subtractor Design III is better compared to the existing design in terms of Quantum Cost, Garbage inputs and Garbage outputs and also in our design the Full Subtraction and Addition function is implemented together as compared to only Subtractor in the existing design. Hence we claim that Design III is better in terms of performance compared to the existing designs.

### 5.3 Simulation Results

Reversible Half, Full Adder/Subtractor and Reversible eight-bit Parallel Binary Adder/Subtractor with Design I, Design II and Design III are implemented using VHDL code and Simulated using Modelsim Simulator. The individual gate functionality is implemented using Behavioral style of Modeling, the overall logic is implemented using Structural style of Modeling and simulation results are shown in shown in Figure 18, 19 and 20.

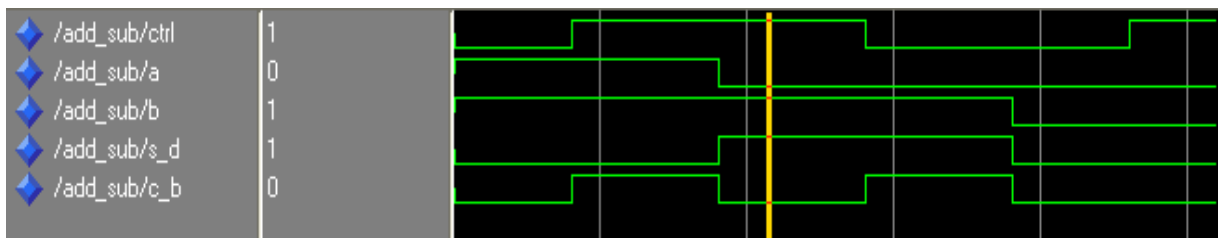

Figure18. Simulation result of Reversible Half Adder/Subtractor

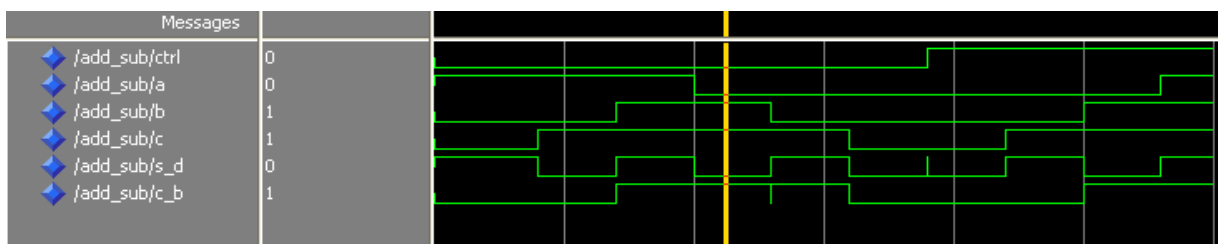

Figure19. Simulation result of Reversible Full Adder/Subtractor

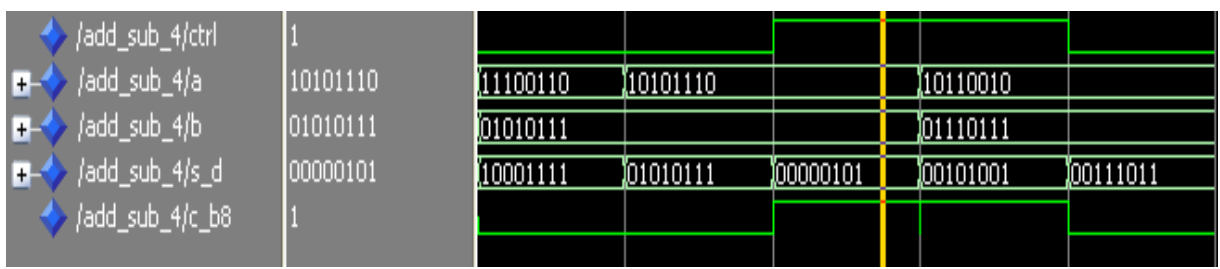

Figure20. Simulation result of Reversible eight-bit Parallel Binary Adder/Subtractor

# 6. Conclusions

The Reversible gates are used to implement Full Adder/Subtractor and Reversible eight-bit Parallel Binary Adder/Subtractor. In this paper, we proposed Reversible eight-bit Parallel Binary Adder/Subtractor unit. The Design I, Design II and Design III are used to implement half and full Adder/Subtractor. The Reversible eight-bit Parallel Binary Adder/Subtractor is built using three designs. The Design III implementation of Reversible eight-bit Parallel Binary Adder/Subtractor has better performance as compared to Design I, Design II and existing design in terms of number of gates used, Garbage inputs/outputs and Quantum Cost, hence can be used for low power applications. The full Adder/Subtractor is implemented in a single unit in our design as compared to only full Subtractor in the existing design [5]. In future, the design can be extended to any number of bits for Parallel Binary Adder/Subtractor unit and also for low power Reversible ALUs, Multipliers and Dividers.

# References

[1] C H Bennett, (1998) "Notes on the History of Reversible Computation", *IBM Journal of Research and Development*, vol. 32, pp. 16-23.

[2] R Landauer, (1961) "Irreversibility and Heat Generation in the Computational Process", *IBM Journal of Research and Development*, vol. 5, no. 3, pp. 183-191.

[3] C H Bennett, (1973) "Logical Reversibility of Computation", *IBM Journal of Research and Development*, vol. 17, no. 6, pp. 525-532.

[4] T Toffoli, (1980) "Reversible Computing", *Technical Memo MIT/LCS/TM-151, MIT Lab for Computer Science*.

[5] H Thapliyal and N Ranganathan, (2009) "Design of Efficient Reversible Binary Subtractors Based on a New Reversible Gate", *IEEE Proceedings of the Computer Society Annual Symposium on VLSI*, pp. 229-234.

[6] H Thapliyal and N Ranganathan, (2010) "Design of Reversible Latches Optimized for Quantum Cost, Delay and Garbage Outputs", *Proceedings of Twenty Third International Conference on VLSI Design*, pp. 235-240.

[7] Lihui Ni, Zhijin Guan, and Wenying Zhu, (2010) "A General Method of Constructing the Reversible Full-Adder", *Third International Symposium on Intelligent Information Technology and Security Informatics*, pp. 109-113.

[8] Irina Hashmi and Hafiz Md. Hasan Babu, (2010) "An Efficient Design of a Reversible Barrel Shifter", *Twenty Third International Conference on VLSI Design*, pp. 93-98.

[9] Robert Wille, Daniel Grobe, D Michael Miller, and Rolf Drechsler, (2009) "Equivalence Checking of Reversible Circuits", *Thirty Ninth International Symposium on Multiple-Valued Logic*, pp. 324-330.

[10] Noor Muhammed Nayeem, Md. Adnan Hossain, Lafifa Jamal, and Hafiz Md. Hasan Babu, (2009) "Efficient Design of Shift Registers using Reversible Logic", *International Conference on Signal Processing Systems*, pp. 474-478.

[11] Majid Mohammadi, Mohammad Eshghi, Majid Haghparast and Abbas Bahrololoom, (2008) "Design and Optimization of Reversible BCD Adder/Subtractor Circuit for Quantum and Nanotechnology Based Systems", *World Applied Sciences Journal,* vol. 4, no. 6, pp. 787-792.

[12] Majid Mohammadi and Mohammad Eshghi, (2008) "Behavioral description of V and $V^+$ gates to Design Quantum Logic Circuits", *Fifth International Multi-Conference on Systems, Signals and Devices*, pp. 1-6.

[13] Rekha K James, Shahana T K, K Poulose Jacob, and Sreela Sasi, (2007) "A New Look at Reversible Logic Implementation of Decimal Adder", *The International Symposium on System-On-Chip*.

[14] Himanshu Thapliyal and A P Vinod, (2006) "Transistor Realization of Reversible TSG Gate and Reversible Adder Architectures", *Proceedings of IEEE Asia Pacific Conference on Circuits and Systems,* pp. 418-421.

[15] Himanshu Thapliyal and M B Srinivas, (2006) "Novel Design and Reversible Logic Synthesis of Multiplexer Based Full Adder and Multipliers", *Forty Eight Midwest Symposium on Circuits and Systems*, vol. 2, pp. 1593 – 1596.

[16] Yvan Van Rentergem and Alexis De Vos, (2005) "Optimal Design of a Reversible Full Adder", *International Journal of Unconventional Computing,* vol. 1, pp. 339 – 355.

[17] Mozammel H A Khan, (2004) "Quantum Realization of Ternary Adder Circuits", *Proceedings of Third International Conference on Electrical and Computer Engineering,* pp. 249-252.

[18] Mozammel H A Khan, (2004) "Quantum Realization of Ternary Toffoli Gate", *Proceedings of Third International Conference on Electrical and Computer Engineering,* pp. 264-266.

[19] Abhinav Agrawal and Niraj K Jha (2004) "Synthesis of Reversible Logic", *Proceedings of the Design, Automation and Test in Europe Conference and Exhibition*.

[20] Pawel Kerntopf, (2002) "Synthesis of Multipurpose Reversible Logic Gates", *Proceedings of the Euromicro Symposium on Digital System Design*, pp. 259-266.

**Authors Short Biography**

**H G Rangaraju** is a Senior Grade Lecturer in the Department of Telecommunication Engineering, Bangalore Institute of Technology, Visvesvaraya Technological University, Belgaum. He obtained his Bachelor degree in Electronics and Communication Engineering from Siddaganga Institute of Technology, Bangalore University and Master degree in Electronics and Communication Engineering from University Visvesvaraya College of Engineering, Bangalore University. He is pursuing Ph.D. in Electronics and Communication Engineering, Visvesvaraya Technological University, Belgaum. His research interest includes VLSI design and Wireless Communication.

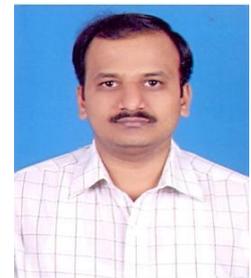

**U Venugopal** obtained his Diploma in Electronics and Communication Engineering from Karnataka Technical Education Board and AMIE digree from Institution of Engineering (India), Calcutta. Presently He is studying M. E. in Electronics and Communication Engineering at University Visveswaraya College of Engineering, Bangalore University. He is working as Engineer in ISRO Satellite Center, Bangalore since 1990 and he has contributed to various satellite programmes of ISRO. He has around three research papers to his credit. His area of interest is VLSI and EMI/EMC.

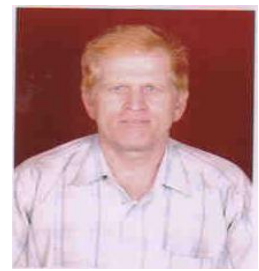

**K N Muralidhara** obtained his BE degree in Electronics and Communication from University of Mysore in 1981. He completed the ME and Ph.D. degrees in 1990 and 1998 respectively from University of Roorkee, Uttaranchal (now known as Indian Institute of Technology, Roorkee). At present, he is working as Professor and Head of the Dept. of Electronics and Communication Engineering, PES College of Engineering, Mandya, Visvesvaraya Technological University. His research interests includes in the areas of Electronic Devices, VLSI, Microprocessor and Microcontroller applications, Embedded systems and Wireless Communication.

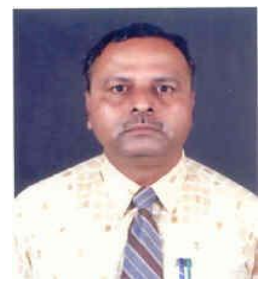

**K B Raja** is an Assistant Professor, Dept. of Electronics and Communication Engineering, University Visvesvaraya College of Engineering, Bangalore University, Bangalore. He obtained his BE and ME in Electronics and Communication Engineering from University Visvesvaraya College of Engineering, Bangalore. He was awarded Ph.D. in Computer Science and Engineering from Bangalore University. He has over 45 research publications in refereed International Journals and Conference Proceedings. His research interests include Image Processing, Biometrics, VLSI Signal Processing, computer networks.

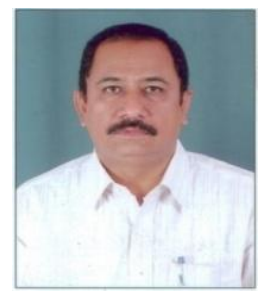